\documentclass[12pt,preprint]{aastex}
\usepackage{tipa}
\usepackage{mathrsfs}

\newcommand{\beq}{\begin{equation}}
\newcommand{\eeq}{\end{equation}}

\shorttitle{Eccentricity Modulation in  GJ436b}
\shortauthors{Tong & Zhou}

\begin{document}

\title{Eccentricity modulation of a close-in planet
by a companion:  application  to GJ 436 system}
\author{Xiao Tong and Ji-Lin Zhou$^\dagger$}
 \affil{Department of Astronomy, Nanjing University, Nanjing
210093, China}  {\footnotetext{$^\dagger$Corresponding author
(email: zhoujl@nju.edu.cn) }} {\footnotetext{ Supported by
National Natural Science Foundation of China (Grant Nos. 10833001
and 10778603), and National Basic Research Program of China
(2007CB4800). }}

\begin{abstract}

GJ 436b is a Neptune-size planet with 23.2 Earth masses in an
elliptical orbit of period 2.64 days and eccentricity 0.16.  With
a  typical tidal dissipation factor ($Q' \sim 10^6$) as that of a
giant planet with convective envelope,
 its orbital circularization timescale under internal tidal dissipation is around $1$ Gyr,
 at least two times less than the stellar age ($>3$~Gyr).
 A plausible mechanism is that the eccentricity of GJ 436b
is modulated by a planetary companion due to their mutual
perturbation. Here we investigate this possibility from the
dynamical viewpoint. A general method is given to predict the
possible locations of the dynamically coupled companions,
including in nearby/distance non-resonant or mean motion resonance
orbits with the first planet.
 Applying the method to GJ 436 system,  we find it is very unlikely that the eccentricity of GJ 436b is maintained at the present location by
  a nearby/distance companion through secular perturbation or
mean motion resonance.  In fact, in all these simulated cases, GJ
436b will undergo eccentricity damp and orbital decay, leaving the
present location within the stellar age.
 However, these results do {\em not} rule out the possible existence of  planet companions in nearby/distance orbits, although
they are not able to maintain the eccentricity of GJ 436b.

\end{abstract}

\keywords{celestial mechanics, methods: analytical, methods: N-body
simulations, stars: individual (GJ 436), planetary systems:
dynamics}

\section{Introduction}

The discovery of the first extrasolar planet around a
pulsar$^{[1]}$, which was quickly followed  by the detection of
first Jupiter-like planet around the star 51 Peg$^{[2]}$, opened a
new era  for planetary science. The planet around 51 Peg is known
as a hot Jupiter: planets with masses comparable to Jupiter's,
 orbits typically within $0.1$ AU and surface temperatures
$\sim 1000$ K$^{[3]}$. To date, more than 310 planets have been
discovered in about 250 planetary
systems\footnote{http://exoplanet.eu/}. With the improvement of
detection precision and the use of various techniques, the minimum
mass of discovered  planets around main sequence stars  is down to
3.3 Earth masses ($M_\oplus$), which is MOA-2007-BLG-192-L b
around a star with mass of $0.06M_\odot$$^{[4]}$.

Among the planets discovered so far, GJ 436b is the only
transiting Neptune-mass planet orbiting an M-type star.  It was
first detected by  radial velocity techniques$^{[5]}$, with its
orbital elements refined later by Maness et al.$^{[6]}$.
 The transiting signals of GJ 436b were first discovered by Gillon et al.$^{[7]}$ and followed
by lots of work$^{[8-13]}$. These observations reveal GJ 436b
with a mass of $23.17M_\oplus$ and a radius of $4.22 R_\oplus$(see
Table 1) .

\begin{table}[!htbp]
\begin{center}
\caption{The parameters of GJ 436 and its planet companion GJ 436.}
\begin{tabular}{lll}
\tableline\tableline Parameter  & Value  & Ref. \\ \tableline
$M_*~(M_\odot)$ & $0.452^{+0.014}_{-0.012}$ &  1 \\
$R_*~(R_\odot)$ & $0.464^{+0.009}_{-0.011}$ & 1 \\
stellar age (Gyr) &    $6^{+4}_{-5}$ &  1 \\ \tableline
$M_p (M_\oplus)$  & $23.17\pm0.079$  & 1  \\
$R_p~(R_\oplus)$ & $4.22^{+0.09}_{-0.10}$ & 1\\
$M_p sin i (M_J)$  & $ 0.0713 \pm 0.006 $  & 2  \\
$a (AU)$ & $0.0285$ & 2  \\
$P (days)$  & $2.64385\pm0.00009$ & 2  \\
$e$ &  $0.16\pm0.019$ & 2 \\
\tableline
\end{tabular}
\end{center} Refs.: 1. \cite{tor07}, 2. \cite{man07}.
\end{table}

One of the most interesting characteristics about  GJ
436b is its significant eccentricity (0.16) in an orbit (2.64
days) very close to the host star.
% According to its mass and radius,
%\citet{Gil07a} think its surface could be the mixture of H/He and
%pure water ice compound according to the mass-diagram  of
%\cite{Fot07}.
Assuming a  tidal dissipation factor ($Q' \sim 10^6$) as that of a
gas giant planet, its orbit circularization timescale under internal tidal dissipation is around $1$ Gyr.
 On the other
hand, the
 fiducial age   of the host star is $6^{+4}_{-5}$~Gyr$^{[11]}$,  and according to observation,
 GJ 436 has low rotation velocity and does not exhibit particular
strong chromospheric activity nor  photometric variability$^{[5]}$,
indicating an age $>3$ Gyr. As  the orbit is not circulated by
planetary tide, either GJ 436b has $Q'>6\times 10^6$, or there is a
planetary companion which  induces a periodic modulation of its
orbital eccentricity.

Considering that radial velocities of GJ 436 reveal a long-term
trend, Maness et al. proposed the presence of a long-period ($\sim
25$ yr) planet companion with mass $\sim 0.27 M_J$ (Jupiter mass)
in an eccentric orbit ($e\sim 0.2$)$^{[6]}$. Recently, Ribas et
al. suggested that the observed radial velocities of the system
are consistent with an additional small, super-Earth planet in the
outer 2:1 mean-motion resonance with GJ 436b$^{[14]}$. Such a
possibility was also studied  from the dynamical
viewpoint$^{[15]}$. More recent inspection of transit data implies
that GJ 436b is perturbed by another planet with mass $\le 12
M_\oplus$ in a non-resonant orbit of 12days (0.08 AU)$^{[16]}$.

In this paper, we exam the possibility of a nearby or distant
undiscovered  planet through dynamical considerations. The key
issue here is to locate the undetected  companion that can both
excite and maintain the eccentricity of GJ 436b. First we present
the dynamical restrictions on the possible locations of the
companion in section 2. Then in section 3 we show the analytical
and numerical results on the eccentricity modulation by a
companion in nearby or distant non-resonance orbits, followed by
the investigation of the possible companion in resonant orbits in
section 4. Conclusions are presented  in section 5.  Although the
method is derived in GJ 436 system, it can be applied to any other
systems in similar situations.

\section{Models and restrictions for planet companion}

Consider a planetary system of a star with mass $M_*$ and two
planets with masses $M_1$ and $M_2$.  For simplicity we assume the
two planets are in coplanar orbits and employ  a general coplanar
three-body model. The orbital elements of the two planets are
denoted by $a_i,e_i,\lambda_i,\varpi_i$, which are semi-major axis,
eccentricity, mean longitude, longitude of periapsis of planet
$i(i=1,2)$, respectively. The index $i$ is labelled so that $M_1$ is the inner
planet with $a_1<a_2$ throughout the evolution. For the present
problem, as it is unlikely to have a planet inside GJ 436b to
modulate its eccentricity, we suppose $M_1$ is GJ 436b and let $M_2$
denote GJ 436c.
Suppose GJ 436b is located in an initially circular orbit with the present semi-major axis,
 we will study the mass range and the location
of the planet companion that excites the eccentricity of  GJ 436b
to the observed value (0.16) due to either secular or resonance interaction.
 In this section  we present  three conditions that restrict the extension of
mass and orbital elements for the undiscovered  companion from (1) Doppler
technique,  (2) Hill stability consideration and (3) tidal
circulation timescale.

\subsection{Observational restriction from Doppler technique}
Radial velocity technique detects the stellar wobbles of orbital
motion in the presence of a planet $M_p$. In terms of the orbital
elements of $M_p$,  the stellar radial velocity is given
as$^{[17]}$:
\begin{equation}
K=\left( \frac{2{\pi}G}{P} \right)^{1/3}\frac{M_p\sin{i}}{(M_{\ast
}+M_p)^{2/3}}\frac{1}{\sqrt{1-e^2}}
\label{K}
\end{equation}
where $P$ is the period of the planet orbit. As $M_p \ll M_*$, the
above equation can be simplified as
\begin{equation}
K \approx 3.0~ {\rm m/s} \left( \frac{P}{10 \rm days} \right)^{-1/3} \left(
\frac{M_p\sin{i}}{10 M_\oplus}\right) \left(  \frac{M_*}{
M_\odot}\right)^{-2/3} \frac{1}{\sqrt{1-e^2}}. \label{K2}
\end{equation}
Due to the perturbation of stellar photosphere, the limit precision
that the Doppler technique can achieve is around 3 m/s$^{[17]}$.
Assume the companion we are to locate is comparable or below this
limit, an undetect planet with stellar radial velocity $K< 3 $ m/s
has maximum mass of,
 \beq
  M_p \sin i \le 5.9  M_\oplus \left( \frac{P}{10 \rm days} \right)^{1/3}(1-e^2)^{2/3}.
\eeq

Using the elements and physical parameters in Table 1, and
assuming $\sin i=0.9784$ by the comparison of GJ 436b's mass from
radial velocity and transit technique, we get the possible
location of companion in the period-mass ($P_{2}$-$M_{2}$) space
in Fig.1.

\begin{figure}[!htbp]
%\vspace*{-1.0 cm}
%\begin{center}
\includegraphics[scale=1.1]{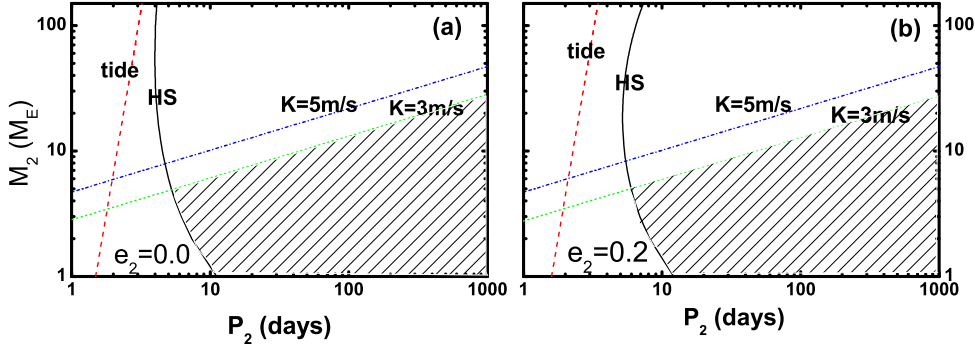}
 \caption{\small Permissible regions (shadow areas)
  of the planet companion in period-mass ($P_2-M_2$) plane  with eccentricity  $e_2=0$ in panel (a) or $e_2=0.2$ in panel (b) under three restrictions:
   (1) radial velocity $K=3$ m/s or 5 m/s, assuming $\sin i=0.9784$(blue and green dot-dash lines, below which the radial velocity is smaller); (2)
   Hill stability (black solid line denoted by `HS', the right part is Hill stable);
  (3) Tidal circulation timescale $\tau_{\rm circ}=6 $Gyr, assuming $Q'_2=10^6$ (red dashed lines, the right part with
  $\tau_{\rm circ}>6$ Gyr).}
   \label{fig1}
%\end{center}
\end{figure}

\subsection{Hill stability condition}

Dynamical stability of the planet system is the necessary
condition for the presence of a second planet.
 One of the practical stability is Hill stability, which requires the ordering of
 two planets unchanged during the history of evolution.
A coplanar, non-resonant,  two-planet system is Hill stable if the
following condition is satisfied$^{[18]}$:
 \beq
H \equiv -\frac{2M_T}{G^2M^3}C^2h~>~ 1+3^{4/3} \frac{M_1M_2}{M_*^{2/3}(M_1+M_2)^{4/3}}+...%-\frac{M_1M_2(11M_1+7M_2)}{3M_*(M_1+M_2)^2}+...,
\label{hill}
 \eeq
where $M_T=M_1+M_2+M_*$ is the total mass of the system, % ($m_1>m_2$ is required here),
 $M=M_1M_2+M_2M_*+M_*M_1$,
$C$ and $h$ are the total angular momentum and energy of the
three-body system, respectively. Since  $M_1, M_2\ll M_*$, by
omitting up to the second order terms in mass ratios of planets to
the star in the expression of total angular momentum and energy,
the left-hand side of equation (\ref{hill}) can be approximated
as, \beq
  H = \frac{1}{(M_1+M_2)^3}\left[ M_1\sqrt{a_1(1-e_1^2)}+M_2\sqrt{a_2(1-e_2^2)}\right]^2 \left[ \frac{M_1}{a_1}+\frac{M_2}{a_2}\right]+ ...
\eeq Denote $\mu_i=M_i/M_*$, $\gamma_i=\sqrt{a_i(1-e_i^2)}$, the
above criterion  for the two planets in Hill stable regions
reads$^{[19]}$, \beq (\mu_1+\mu_2)^{-3}(\mu_1\gamma_1+\mu_2
\gamma_2)^2\left(\frac{\mu_1}{a_1}+\frac{\mu_2}{a_2}\right)>
1+3^{4/3} \mu_1\mu_2 (\mu_1+\mu_2)^{-4/3}. \label{HS}\eeq
 The Hill stability
region  by Eq. (\ref{HS}) is also plotted in Fig.1.

\subsection{Restriction from tidal circulation timescale}

A close-in planet produces tidal bulges on the stellar surface,
causing energy dissipation on the star and angular momentum
exchanges between the stellar spin and planetary orbital motion.
 Meanwhile the star also generates tidal dissipation on the planet, resulting in an
 eccentricity damping and  orbital decay$^{[20,21]}$. In the ideal case that both the stellar and
planetary spins are aligned with the orbit, the secular evolution
rate of the eccentricity can be expressed as follows $^{[22-24]}$,
\begin{equation}
\dot{e}=g_p+g_*
\label{edot}
\end{equation}
\begin{equation}
g_{p,*}=\left( \frac{81}{2} \frac{ne}{Q'_{p,*}}\right) \left(
\frac{M_{*,p}}{M_{p,*}}\right) \left( \frac{R_{p,*}}{a}\right)^5
\left[ -f_3(e)+\frac{11}{18}f_4(e) \left(
\frac{\Omega_{p,*}}{n}\right)  \right], \label{gp*}
\end{equation}
where  $g_{p}$ and $g_{*}$  are contributions from the planet and
star, respectively,  $n$ is the mean velocity of orbital motion,
$M_{*,p}, Q'_{*,p}, R_{*,p}, \Omega_{*,P}$ are  the masses, the
effective tidal dissipation factors, radii and spin rates
 of the star and the planet, respectively.  Some functions  of eccentricity used here and later are:
\beq
\begin{array}{l}
 %f_2(e)=(1+\frac{15}{2}e^2+\frac{45}{8}e^4+\frac{5}{16}e^6)/(1-e^2)^{6}, \\
 f_3(e)=(1+\frac{15}{4}e^2+\frac{15}{8}e^4+\frac{5}{64}e^6)/(1-e^2)^{13/2},\\
f_4(e)=(1+\frac{3}{2}e^2+\frac{1}{8}e^4)/(1-e^2)^{5}, \\
%f_5(e)=(1+3e^2+\frac{3}{8}e^4)/(1-e^2)^{9/2}, \\
f_6(e)= (1+\frac{15}{7}e^2+\frac{67}{14}e^4+\frac{85}{32}e^6+
\frac{255}{448}e^8+\frac{25}{1792}e^{10})/(1+3e^2+\frac38 e^4), \\
f_7(e)=(1+\frac{45}{14}e^2+8e^4+\frac{685}{224}e^6+
\frac{255}{448}e^8+\frac{25}{1792}e^{10})/(1+3e^2+\frac38 e^4).
\end{array}
\eeq
 For close-in planets with tidal dissipation
factor $Q'_p \le 10^6$, dissipation in planets dominates$^{[25]}$.
Omitting contribution from the star in Eq. (\ref{edot}) and
assuming the planet spin has reached the synchronization
equilibrium ($\Omega_p \sim n$), the timescale of orbital
circularization ($\tau_{\rm circ}=e/\dot{e}$) induced by planetary
tidal dissipation is given as$^{[25]}$,
\begin{equation}
%\begin{array}{ll}
 \tau_{\rm circ} %& =\frac{(1-e^2)^{13/2}}{f_7(e)} \left( \frac{4Q'_p}{63n}  \right) \left(\frac{M_p}{M_*} \right)\left(\frac{a}{R_p} \right)^5 \\
% &
\simeq 3.6 \frac{(1-e^2)^{13/2}}{f_7(e)} \left(\frac{Q'_p}{10^6} \right)
\left(\frac{M_p}{M_J} \right) \left( \frac{M_*}{M_\odot} \right)^{2/3}
 \left( \frac{P}{\rm 1~ day}\right)^{13/3}    \left(\frac{R_p}{R_J}\right)^{-5} {\rm Myr}. % \\
 \label{tcir}
%\end{array}
\end{equation}
And the associate timescale of orbital decay ($\tau_{\rm decay}=
a/\dot{a} $) in elliptical orbits is \beq \tau_{\rm decay}
=\frac{(1-e^2)f_7(e)}{2e^2f_6(e)} \tau_{\rm circ}. \eeq For GJ
436b, with the elements and physical parameters in Table 1 and
assuming $Q'=10^6$, the circularization  timescale of GJ 436b is
1.0 Gyr, around five times less than the fiducial stellar age(6
Gyr), and $\tau_{\rm decay}=19 $ Gyr for GJ 436b at the present
location with $e=0.16$. So some mechanisms are needed to maintain
its eccentricity during the evolution.

\section{Planet Companions in Non-Resonant Orbits}

Secular perturbations between two planets in non-resonance orbits
exchange their angular momentum, thus modulate their
eccentricities, leaving their semi-major axes almost unchanged.
When tidal dissipation is present on either of the planet,
eccentricity modulation is effective only when the  timescale of
secular perturbation ($\tau_{\rm sec}$)  is significantly shorter
than that of the circularization ($\tau_{\rm circ}$). With an
octopole Legendre expansion in ratio $\alpha=a_1/a_2$, Mardling
%\cite{mar07}
 derived the period of secular perturbation at the
limit of  $e_1\ll 1$, including the effect of general relativity$^{[26]}$,
\begin{equation}
 \tau_{\rm sec} = \frac{4}{3}P_1 \alpha^{-3} \left(  \frac{M_2}{M_*}\right)^{-1}\varepsilon_2^3
   \left [ 1-\sqrt{\alpha}\varepsilon_2^{-1} \left(  \frac{M_1}{M_2}\right) +\gamma\varepsilon_2^3 \right]^{-1},
   \label{tsec}
\end{equation}
where $P_1$ is the mass of first planet ($M_1$),
$\varepsilon_2=\sqrt{1-e_2^2}$, $
\gamma=4\alpha^{-3}(n_1a_1/c)^2(M_{\ast }/M_2)$ is the ratio of
general relativity to companion perturbation on periapsis
precession of $M_1$, with $n_1$ the  mean motion of $M_1$ and $c$
the speed of light. For a companion in a nearby orbit,
 $\tau_{\rm sec} \ll  \tau_{\rm circ}$ is easily fulfilled, as
$\gamma  \ll 1 $ and $\alpha$ is moderate. For a giant companion
in a distance orbit, $\gamma \gg 1$, thus  Eq. (\ref{tsec}) can be
simplified as $    \tau_{\rm sec} \approx P_1/3 /(n_1a_1/c)^2 =1.5
\times 10^4 $ yrs, which is independent of the planet companion and
is much shorter than the circularization timescales of both
planets.

%Under tidal dissipation, the eccentricity will evolves into libration  around an equilibrium values within a relatively short period.
Under secular perturbation of $M_2$, the maximum eccentricity of
$M_1$ that can be achieved from an initial circular orbit
$(e_{10}=0)$ is$^{[26]}$,
\begin{equation}
e_{\rm 1max}=\frac{5}{2} \alpha e_2\varepsilon_2^{-2}
\left|1-\sqrt{\alpha}\varepsilon_2^{-1} \left(
\frac{M_1}{M_2}\right)+\gamma\varepsilon_2^3\right|^{-1}.
\label{e1max}
\end{equation}
This equation can be used to locate a global approximate region of the
planet companion in either  nearby or distance orbits,
 while general three-body simulations should be performed to give a precise
location.

Eq.(\ref{e1max}) provides {\em a mechanism} that can excite
$e_1$ to a moderate value at $M_1 > M_2$.  In fact, from the
expression of $e_{\rm 1max}$, there exists  a singularity
 at $\alpha_{\rm crit}$  when the denominate of
Eq. (\ref{e1max}) is zero so that $e_{\rm 1max}$ tends to infinity
when  $M_1 > M_2$ . In reality, as $e_{\rm 1max}$ becomes large
enough, the above approximation $e_1 \ll 1$ is no more valid and
we should resort to numerical simulations.
 Fig.2a shows the maximum eccentricity ($e_{\rm 1max}$) as a function of companion's period and mass ($P_{2}$-$M_{2}$)
  derived from Eq.(\ref{e1max}) with $e_{20}=0.2$.
To verify this, we perform some three-body simulations with $M_1$
in initial circular orbits,  and obtain $e_{\rm 1max}$
 under the modulation of $M_2$ at different locations.
 Fig.2b shows the results by three-body simulations.  The critical locations ($\alpha_{\rm crit}$) when the singularity occurs
 in Eq.(\ref{e1max})
  are also plotted in the curves with the asterisks, showing a roughly good agreement between the
 analytical and numerical results.   According to Fig.2, it is possible to excite  $e_{\rm 1max}=0.16$ in close
 orbits with $M_2 \ge 10 M_\oplus$ and $e_{2}=0.2$. For
 gas giant companion, only nearby orbit is possible. We investigate these situations in detail as follows.

\begin{figure}[!htbp]
\vspace*{-1.0 cm}
\includegraphics[scale=1]{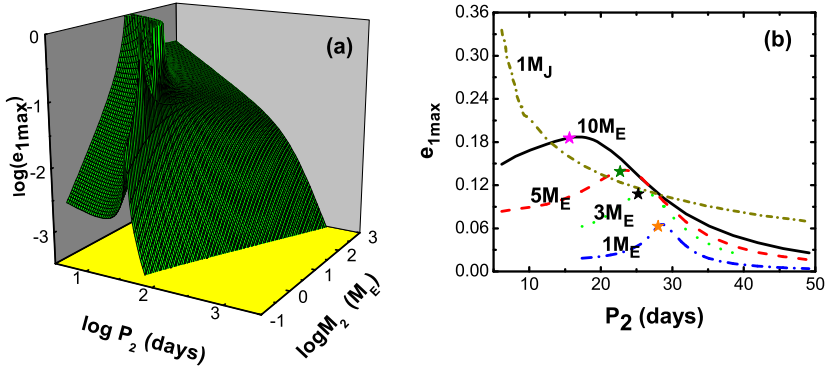}
 \caption{\small Maximum eccentricity of $M_1$ that can be excited at the present location
   by a planet companion  with initial eccentricity $e_{2}=0.2$. Panel (a): 3-D plot of
   $e_{\rm 1max}$ in the plane of  $P_2-M_2$ from theoretical Eq. (\ref{e1max}).
   Panel (b): $e_{\rm 1max}$ obtained from numerical simulations of a general three-body model.
    The asterisks denote the singularities from Eq. (\ref{e1max}).}
   \label{fig2}
\end{figure}

\noindent{\em Nearby Orbits.} We perform extensive three-body
simulations, including general relativity effect, on
 the initial $P_2-e_2$ plane, with a nearby companion mass
of $5 M_\oplus $ or $10 M_\oplus$. The results are shown in
Fig.3a,b. The shadow regions in Fig.3a,b  are the most possible
locations of the companion that can maintain $e_1=0.16$ by secular
perturbation, combined with the three restrictions present in
section 2. As two examples, we present the evolution of  two
orbits from the permissible regions under tidal dissipation (with
illustrative $Q'_1=Q'_2=100$) in Fig.3c,d. The eccentricity of GJ
436b  can be excited and maintained (with a periodic modulation)
to 0.16 only within $10^5$ years. Considering the linear
dependence of tidal force on $Q'_i$, (i=1,2), these simulations
indicate that $e_{\rm 1max}  \sim 0.16$ can  be maintained {\em
only} for 1 Gyr provide $Q'_i =  10^6$.

\begin{figure}[!htbp]
%\vspace*{-1.0 cm}
%\begin{center}
\includegraphics[scale=1.2]{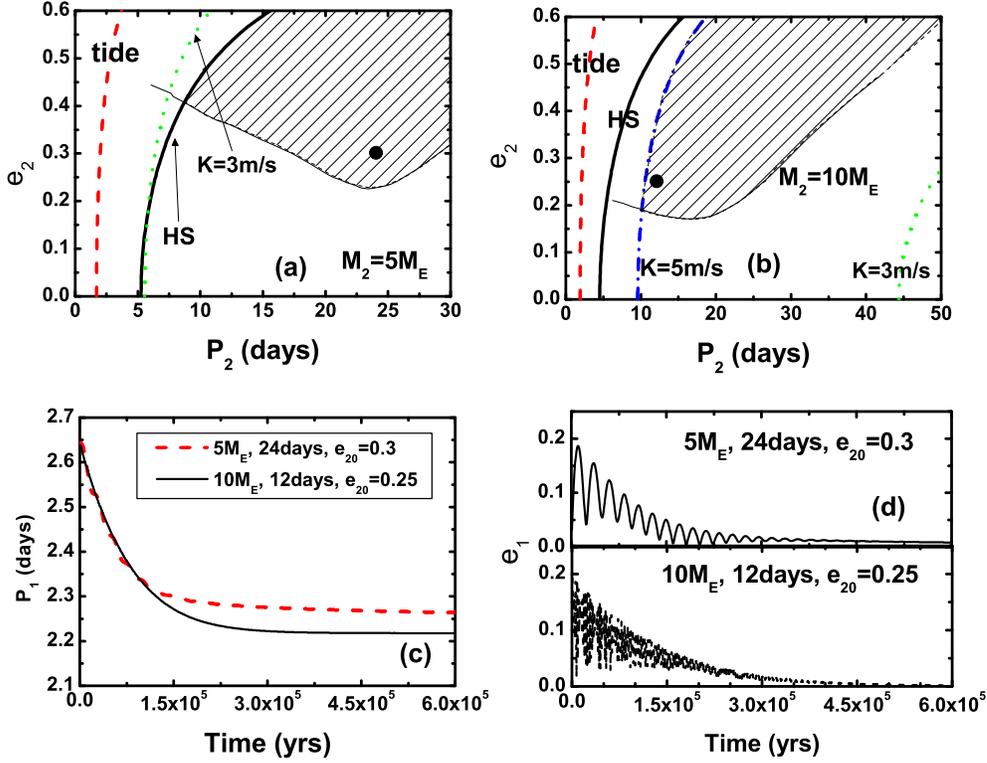}
 \caption{\small Permissible regions and orbital evolution for the companion $M_2$ in nearby orbits  that can excite $e_{1}=0.16$ at the present location.
 Panels (a) and (b) are permissible regions (shadow areas) in $P_2-e_2$ plane obtained  from general  three-body simulations for companion mass
 $5M_\oplus$ and $10M_\oplus$, respectively.
 The boundary lines  of the three restrictions ($K$= 3 m/s or 5 m/s; Hill stability denoted by HS, and
 $\tau_{\rm circ}=6 $Gyr with $Q'_2= 10^6$)  are also
 plotted.  Evolution of $P_1,e_1$  for two orbits (denoted by filled circles) in
panels (a) and (b) are shown in panels  (c) and (d), respectively,
with initial parameters shown in the labels. Tidal dissipation
factors of two planets are taken $Q'_2=100$ in panels (c) and (d),
 with density of $M_2$ taken as $\rho_2=3$ g cm$^{-3}$.}
   \label{fig3}
%\end{center}
\end{figure}

\noindent{\em Distant Orbits.} According to  equation
(\ref{e1max}) and numerical simulation (Fig.2b),  $e_{\rm 1max}$
is small unless $M_2$ is in a highly eccentric orbit. Fig. 4a
plots the  region in $P_2-e_2$ plane that a  companion can
generate a maximum eccentricity $e_{\rm 1max}=0.16$. They are
calculated from equation (\ref{e1max}) and confirmed by full
3-body simulations. Thus it is almost impossible for a companion
in orbit of $P_2\sim 1$ yr to produce $e_{\rm 1max}=0.16$. The two
possible locations of $M_2$ suggested by Maness et al.$^{[6]}$ are
also investigated and plotted in Fig.4d, which shows they can only
excite negligible eccentricities of $M_1$.

\begin{figure}[!htbp]
%\vspace*{-1.0 cm}
%\begin{center}
\includegraphics[scale=0.8]{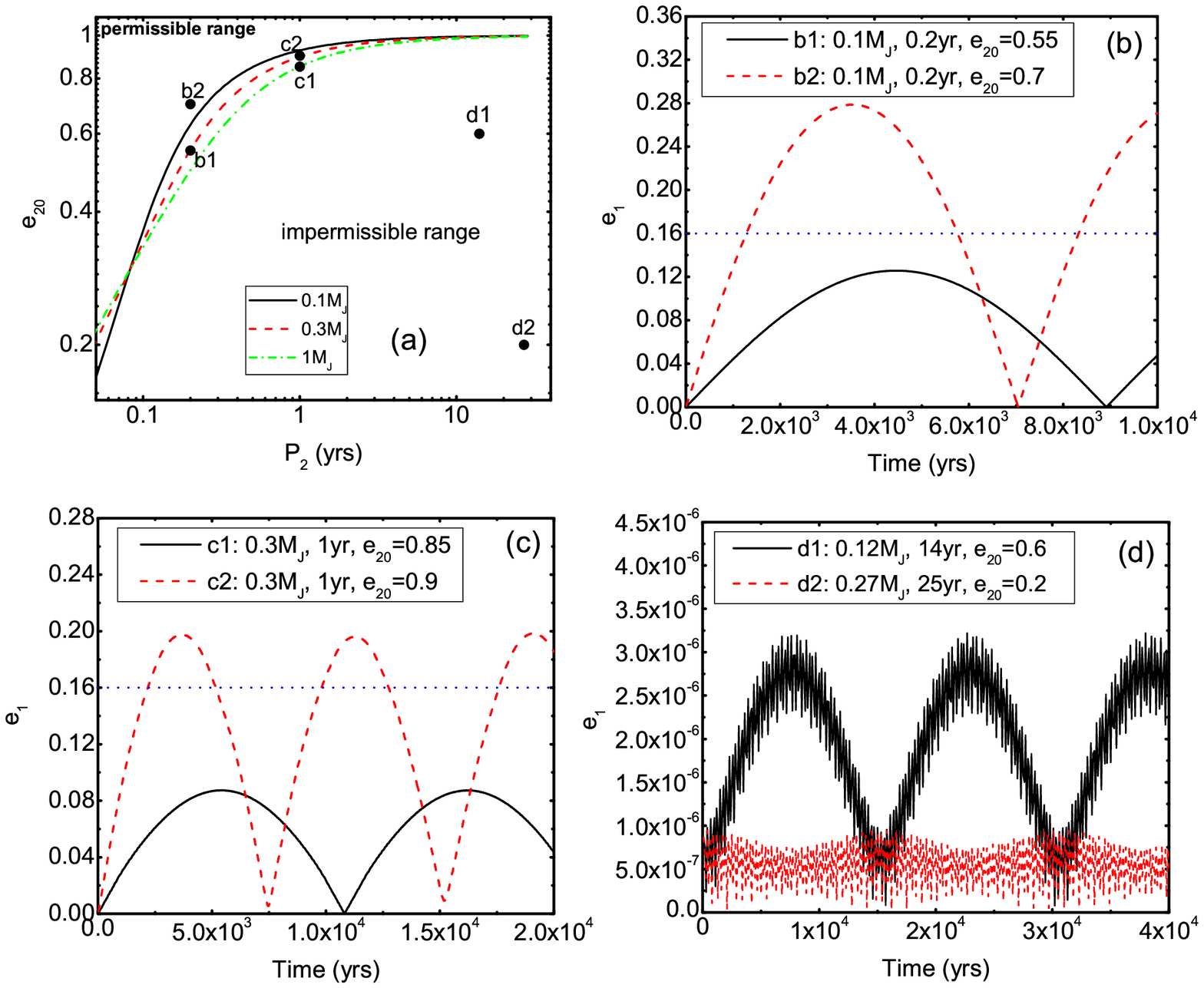}
 \caption{\small Permissible region and orbital evolution for distance giant companion $M_2$ that can  excite $e_{1}=0.16$ at the present location. Panel (a): Minimum
 eccentricities of companion with different mass $M_2=0.1 M_J, 0.3M_J, 1M_J$ from Eq. (\ref{e1max}) and are confirmed by general three-body simulations.
 Panels (b-d): Evolution of $e_1$ of six orbits notated in panel (a) by three-body simulations, with
 mass and initial elements shown in the labels. Orbits in panel (d) are proposed by Maness et al.$^{[6]}$.}
   \label{fig4}
%\end{center}
\end{figure}

\section{Planet Companion in Resonant Orbits}

For a conservative coplanar two-planet system, the motion of two
planets can be described by a Hamiltonian system with four degrees
of freedom, which  is non-integrable. However, near a generic
$(p+q)/p ~(q\neq 0)$ mean-motion resonance,  the degrees of freedom
of the system are reduced from four to two by averaging
technique$^{[27]}$. Below we will show that, the minimum initial
eccentricity  of unseen planet $M_2$ can be deduced approximately
from the conservation of total orbital energy and angular momentum,
with the help of the averaged Hamiltonian function.

%\subsection{Hamiltonian in Mean Motion Resonance}

Adopt the following planar canonical  variables$^{[27,28]}$,
\begin{equation}
\begin{array}{ll}
\lambda_1, & J_1=L_1+s(I_1+I_2),  \\
\lambda_2, & J_2=L_2-(1+s)(I_1+I_2), \\
\sigma_1=(1+s)\lambda_2-s\lambda_1-\varpi_1, & I_1=L_1(1-\sqrt{1-e_1^2}), \\
\sigma_2=(1+s)\lambda_2-s\lambda_1-\varpi_2, & I_2=L_2(1-\sqrt{1-e_2^2}).
\end{array} \label{can}
\end{equation}
where $\lambda_i,\varpi_i$ are the longitude of mean motion,
longitude of periapsis of $M_i~(i=1,2)$, respectively, and
\begin{equation}
\begin{array}{ll}
s=p/q, &
 L_i=M_i'\sqrt{\mu_i a_i},~(i=1,2) \\
\mu_i=G(M_*+M_i),& M_i'=\frac{M_iM_*}{M_i+M_*}.
\end{array}
\end{equation}
with $G$  the gravitational constant. The Hamiltonian $H$ of the system can be expressed as:
\begin{equation}
H=H_0+H_1,
\label{ham}
\end{equation}
where the first term corresponds to the two-body contribution given by:
\begin{equation}
H_0=-\sum_{i=1}^2\frac{\mu_i^2{M_i^{\prime }}^3}{2L_i^2}.
\end{equation}
The second term, $H_1$, is the disturbing function. Up to the first
order in the masses, it has the following expression$^{[28]}$:
\begin{equation}
H_1=-G\frac{M_1M_2}{\Delta
}+\frac{M_1M_2}{M_*}(\dot{x_1}\dot{x_2}+\dot{y_1}\dot{y_2}+\dot{z_1}\dot{z_2})
\end{equation}
where $\Delta=(r_1^2+r_2^2-2r_1r_2\cos{S})^{1/2}$, and for the
coplanar three-body system, $S=f_1-f_2+\varpi_1-\varpi_2$ with $f_i$
the true anomaly of the orbits $m_i~(i=1,2)$. In terms of the
elements on (\ref{can}), all periodic  terms in the Hamiltonian
(\ref{ham}) contain only three independent angular variables
$\sigma_1,\sigma_2,\lambda_1-\lambda_2$, thus the system is three
degrees of freedom$^{[27]}$. The canonical moment conjugate
$\lambda_1+\lambda_2$ is an integral of motion, namely $J_1+J_2={\rm
const}$. By averaging the synodic angle $Q=(\lambda_1-\lambda_2)/q$,
we obtain an averaged system with Hamiltonian function,
\begin{equation}
\bar{H}=H_0+\frac{1}{2\pi}\int_0^{2\pi}H_1dQ
\label{aveH}
\end{equation}
In practice, the above averaged Hamiltonian can be obtained only
numerically.

The averaged system with  Hamiltonian (\ref{aveH}) is of two
degrees of freedom, with the energy being the only integral. To show all
possible solutions in $(e_1,e_2)$ space for all possible phase
angles $\sigma_1,\sigma_2$ is impossible. So we fix only the
symmetric resonance period orbits with initial $\sigma_1,\sigma_2$
being set at either 0 or $\pi$. Fig.5a shows the
energy level curves of Hamiltonian at $10M_{\oplus}$ from
equation (\ref{aveH}) on the ($e_1,e_2$) plane.   Based on the
contour lines (or the averaged Hamiltonian function, equivalently) and
fixed on symmetric period orbits, we derive the minimum initial eccentricity of the
companion, $e_{\rm 2min}$, with which $M_1$ can evolve to $e_1=0.16$,
as a function of companion mass (Fig.5b). Interestingly,
$e_{\rm 2min}$ has a power-law dependence on the mass ratio $M_1/M_2$, with
an approximation relation:
\begin{equation}
e_{\rm 2min}\approx 0.14 (\frac{M_1}{M_2})^{1/2}
    \label{e2min}
\end{equation}
The relation holds for the three major resonances 2:1,3:1 and 5:2,
and is independent of the location of $M_1$.

\begin{figure}[!htbp]
\vspace*{0 cm}
\begin{center}
\includegraphics[width=5.0in]{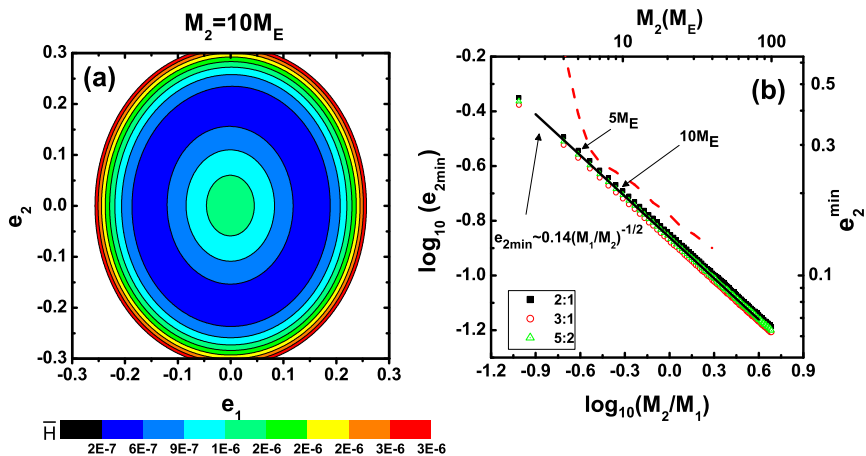}
 \caption{\small Panel (a): Constant-energy contour of the average Hamiltonian (\ref{aveH}) with $M_2=10M_{\oplus}$.
 Panel (b): Minimum eccentricity of $M_2$ that can excite $e_{1}=0.16$ according to the energy level curves at 2:1,3:1 and 5:2 resonances.
 The red dashed curve is calculated by three-body simulations for the 2:1 resonance.}
   \label{fig5}
\end{center}
\end{figure}

The above relation can be deduced from the conservation of total
angular momentum,
$J_1+J_2=L_1\sqrt{1-e_1^2}+L_2\sqrt{1-e_2^2}={\rm
const}$(independent of $s$). In fact, in an ideal situation such
that  $e_{\rm 1max}$ ($e_{\rm 2min}$) occurs at $e_2=0$($e_1=0$,
respectively), the conservation of total angular momentum
requires, $L_{10}\sqrt{1-e_{\rm
1max}^2}+L_{20}=L_{10}+L_{20}\sqrt{1-e_{\rm 2min}^2}$,  which
gives approximately, $e_{\rm 2min}=\sqrt{L_{10}/L_{20}}e_{\rm
1max} \approx (M_1/M_2)^{1/2}(n_2/n_1)^{1/6}e_{\rm 1max}$, where
$L_{10},L_{20}$ are the corresponding elements evaluated at the
resonance center. It shows that $e_{\rm 2min}$ depends weakly on
the mean motion ratio.   At the 2:1 resonance, let $e_{\rm
1max}=0.16$, we derive the approximate formula (\ref{e2min}).

From the above derivation of equation (\ref{e2min}), we can see that,
the specific resonance structure, which would be very complicated
in a general three-body model, is not considered. So the
 relation (\ref{e2min}) holds approximately only, and for a real $e_{\rm 2
 min}$, we shall resort to  numerical simulations. Fig.5b plots also the
 results from the  three-body simulations including the general relativity
 effect. The discrepancy between the relation and the simulation is large especially when
 $M_2$ is small.

Fig.6 shows the evolution of two typical orbits
  in 2:1 mean motion resonance with $M_1$ for a companion $M_2=5M_{\oplus}$, as proposed by Ribas et al.$^{[14]}$.
  The eccentricity they proposed is 0.2, below the value of $e_{\rm 2min}$ plotted in Fig.5b.
 As we can see, $e_1$ can not be excited to 0.16(Fig.6b).
 For the higher $e_{20}=0.40$ case, it can excite to $e_1=0.16$ initially,
 but the eccentricity is damped soon. In both cases the orbit of GJ 436 decays to inner orbits.
Other mean motion resonances show similar results, indicating that it is {\em impossible} to maintain $e_1=0.16$ by a resonant
companion.

\begin{figure}[!htbp]
\vspace*{0 cm}
\begin{center}
\includegraphics[width=5.0in]{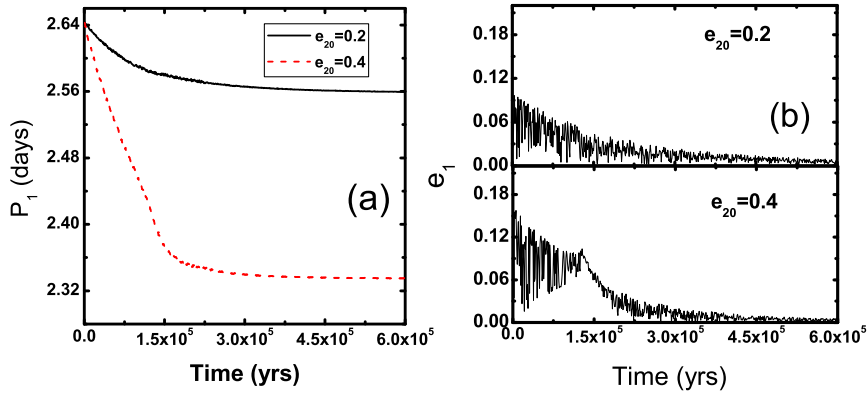}
 \caption{\small Evolution of $P_1, e_1$ of two orbits initially in 2:1 resonance with $M_1$ by three-body simulations. Companion mass
 $M_2=5M_\oplus$ and density $\rho_2=3$ g cm$^{-3}$, initial eccentricities of $M_2$ are $0.2$ and $0.4$.
   The orbit with $e_{20}=0.2$  was proposed by Ribas et
al.$^{[14]}$.
  Tidal dissipation factors of two planets are taken as $Q'=100$.}
   \label{fig6}
\end{center}
\end{figure}

\section{Conclusions}

Perturbation from a companion $M_2$ seems to be the most
plausible approach to maintain the  moderate eccentricity  of GJ 436b in a close-in orbit.
  In this paper, we study
extensively all different locations of the planet companion in the
following three  situations:

{\em (i) In nearby orbits}. The eccentricity of GJ 436b can be
excited to 0.16 with a broad range of planet mass (above few
Earth-masses). An interesting case is the second planet can have a
mass much smaller than $M_1$. However, since the orbital decay
time is  short ($\sim 20 $Gyr) for GJ 436b at the present location
with the observed eccentricity, significant orbital decay $(\sim
25\%)$ is expect so that GJ 436b would be in a much closer orbit,
and the eccentricity of GJ 436b would be damped within the stellar
age. Thus the eccentricity of GJ 436b can  {\em not} be maintained
by a nearby companion. On the other hand, one can {\em not} rule
out the possibility that companions exist in nearby orbits of GJ
436b as long as they are in stable orbits, e.g., the possible
existence of a planet with $\le 12 M_\oplus$ in an orbit of 12days
from  transit data$^{[16]}$.  However, the companion, if exists,
does not account for the eccentricity of GJ 436b.

 {\em (ii) Distance Orbits}.
Distance companions on moderate eccentric orbits can {\em not}
excite and maintain  the significant eccentricity of GJ 436b.
However, the presence of giant planets in extended orbits ($\sim $
few AU) is also possible, although they can {\em not} modulate the
eccentricity of close-in planets. For example, there may exist a
long-period ($\sim 25$ yr) planet companion with mass $\sim 0.27
M_J$ inferred from radial velocities of  GJ 436 $^{[6]}$.

{\em (iii) Resonance Orbits}. Although a border range of companion
mass can excite the eccentricity of $M_1$ to  a moderate value,
similar to situation ({\em i}), significant orbital decay would
occur so that the two planets will leave 2:1 resonance soon, if GJ
436b has a normal dissipation factor ($Q'\sim 10^6$). In this case,
the presence of a companion in the outer 2:1 resonance  with GJ 436b
is unlikely. However, if we have to confess that GJ 436b has a
extremely high dissipation factor ($Q'> 6 \times 10^6$), the
existence of companions in resonance orbits can {\em not} be ruled
out in this extreme case, e.g., the planet ($\sim 5 M_\oplus$) in
the outer  2:1 resonance place proposed by  Ribas et al.$^{[14]}$.
However, even it exists, it does not account for the eccentricity of
GJ 436b.

Based on the extensive investigations in this paper, we think
$e_1=0.16$ of GJ 436b can not be maintained by a companion  in
either nearby or distance orbits through secular perturbation or
mean motion resonance. Thus the maintaining of its eccentricity
remains a challenge problem, unless GJ 436b has a extremely high
dissipation factor ($Q'> 6 \times 10^6$).

\acknowledgments

{\bf Acknowledgments}.
 We  thanks Dr. Gregory Laughlin for useful discussions.


\begin{thebibliography}{99}

\bibitem[Wolszczan
\& Frail(1992)]{WF92} 1 Wolszczan, A., \& Frail, D.~A., A planetary
system around the millisecond pulsar PSR1257 + 12, 1992, \nat, 355,
145

\bibitem[Mayor
\& Queloz(1995)]{May95} 2 Mayor, M., \& Queloz, D., A Jupiter-mass
companion to a Solar-type star, 1995, \nat, 378, 355

\bibitem[Udry \& Santos(2007)]{US07} 3 Udry, S., \& Santos, N.~C., Statistical properties of exoplanets, 2007, \araa, 45, 397

\bibitem[Bennett et al.(2008)]{ben08} 4 Bennett, D. P., Bond, I. A., Udalski, A., et
al., A low-mass planet with a possible sub-stellar-mass host in
microlensing event MOA-2007-BLG-192, 2008, \apj, 684, 663

\bibitem[Butler et al.(2004)]{But04} 5 Butler, R.~P.,
Vogt, S.~S., Marcy, G.~W., et al., A Neptune-mass planet orbiting
the nearby M dwarf GJ 436, 2004, \apj, 617, 580

\bibitem[Maness et al.(2007)]{man07} 6 Maness, H.~L., Marcy,
G.~W., Ford, E.~B., et al., The M dwarf GJ 436 and its Neptune-mass
planet, 2007, \pasp, 119, 90

\bibitem[Gillon et al.(2007a)]{Gil07a} 7 Gillon, M., Demory, B.-O.,
 Barman, T., et al., Accurate spitzer infrared radius measurement for the hot Neptune GJ 436b, 2007a, \aap, 471, L51

\bibitem[Gillon et al.(2007b)]{Gil07b} 8 Gillon, M., Pont, F.,
 Demory, B.-O., et al., Detection of transits of the nearby hot Neptune GJ 436b, 2007b, \aap, 472, L13

\bibitem[Deming et al.(2007)]{demi07} 9 Deming, D., Harrington,
J., Laughlin, G., et al, Spitzer transit and secondary eclipse
photometry of GJ 436b, 2007, \apjl, 667, L199

\bibitem[Demory et al.(2007)]{demo07} 10 Demory, B.-O., Gillon, M., Barman, T.,
et al., Characterization of the hot Neptune GJ 436b with spitzer and
ground-based observations 2007, \aap, 475, 1125

\bibitem[Torres(2007)]{tor07} 11 Torres, G., The transiting exoplanet host star GJ 436: A test of stellar evolution models in the lower main sequence, and revised planetary parameters, 2007, \apjl, 671, L65

\bibitem[Bean et al.(2008)]{bean08} 12 Bean, J. L., Benedict, G. F., Charbonneau, D.,
et al., A Hubble space telescope transit light curve for GJ 436b,
2008, \aap, 486, 1039

\bibitem[Shporer et al.(2008)]{Shp08} 13 Shporer, A., Mazeh, T.,
Winn, J.~N., et al., Photometric follow-up observations of the
transiting Neptune-mass planet GJ 436b, 2008, arXiv:0805.3915

\bibitem[Ribas et al.(2008)]{rib08a} 14 Ribas, I., Font-Ribera, A., \& Beaulieu, J.-P., A $\sim$ 5 $M_{\oplus}$ super-earth orbiting GJ 436? The power of near-grazing transits, 2008, \apjl, 677,
L59

\bibitem[Mardling(2008)]{mar08} 15 Mardling, R.~A., On the long-term tidal evolution of GJ 436b in the presence of a resonant companion, 2008, arXiv:0805.1928

\bibitem[Coughlin et al.(2008)]{cou08} 16 Coughlin, J.~L., Stringfellow, G.~S., Becker, A.~C.,
et al., New observations and a possible detection of parameter
variations in the transits of Gliese 436b, 2008, arXiv:0809.1664

\bibitem[Marcy \& Butler(1998)]{marcy98} 17 Marcy, G.~W., \& Butler, R.~P., Detection of extrasolar giant planets, 1998,
\araa, 36, 57

\bibitem[Marchal \& Bozis(1982)]{MB82} 18 Marchal, C., \& Bozis, G., Hill stability and distance curves for the general three-body problem, 1982, Cele. Mech., 26, 311

\bibitem[Gladman(1993)]{gla93} 19 Gladman, B., Dynamics of systems of two close planets, 1993, Icarus, 106,247

\bibitem[Goldreich \& Soter(1966)]{GS66} 20 Goldreich, P., \& Soter, S., Q in the solar system, 1966, Icarus, 5, 375

\bibitem[Murray \& Dermott(1999)]{MD99} 21 Murray, C.~D., \& Dermott, S.~F.\ 1999,
Solar system dynamics, Cambridge, Cambridge unversity Press, C.~D.,
1999

\bibitem[Eggleton et al.(1998)]{egg98} 22 Eggleton, P.~P.,
Kiseleva, L~G., \& Hut, P., The equilibrium tide model for tidal
friction, 1998, \apj, 499, 853

\bibitem[Mardling \& Lin(2002)]{ML02} 23 Mardling, R.~A., \& Lin, D.~N.~C., Calculating the tidal, spin, and dynamical evolution of extrasolar planetary systems, 2002, \apj,
573, 829

\bibitem[Dobbs-Dixon et al.(2004)]{Dob04} 24 Dobbs-Dixon, I.,
Lin, D.~N.~C., \& Mardling, R.~A., Spin-orbit evolution of
short-period planets, 2004, \apj, 610, 464

\bibitem[Zhou \& Lin(2008)]{ZL08} 25 Zhou, J.-L., \& Lin, D.~N.~C., Migration and final location of hot super earths in the presence of gas giants, in {\it
Exoplanets: Detection, Formation and Dynamics}, eds: Sun,Y.-S.,
Ferraz-Mello,S., Zhou,J.-L., Proc. of IAU Symp. 249, Cambridge,
Cambridge unversity Press, 2008, 285-291

\bibitem[Mardling(2007)]{mar07} 26 Mardling, R.~A., Long-term tidal evolution of short-period planets with companions, 2007,
\mnras, 382, 1768

\bibitem[Beaug{\'e} \& Michtchenko(2003)]{BM03} 27 Beaug{\'e}, C., \& Michtchenko,
T.~A., Modelling the high-eccentricity planetary three-body problem.
Application to the GJ876 planetary system, 2003, \mnras, 341, 760

\bibitem[Laskar(1991)]{las91} 28 Laskar, J., NATO Advanced
Study Institute on Predictability, Stability, and Chaos in N-Body
Dynamical Systems (A.E.Roy, ed.), Plenum Press, New York, 1991, 93
- 114,



\end{thebibliography}
\end{document}